\documentclass[twocolumn]{aa}
\usepackage{graphicx}
\usepackage{natbib}
\bibpunct{(}{)}{;}{a}{}{,}
\usepackage{txfonts}
\begin{document}
   \title{The XMM-Newton observation of GRB 040106 : evidence for an afterglow in a wind environment.}
   \author{B. Gendre
          \inst{1}
       L. Piro
      \inst{1}
          \and
          M. De Pasquale\inst{1}
          }

   \offprints{B. Gendre}

   \institute{Istituto di Astrofisica Spaziale e Fisica Cosmica/CNR, via fosso del cavaliere, 100, 00133, Roma, Italy\\
              \email{gendre@rm.iasf.cnr.it, piro@rm.iasf.cnr.it, depasq@rm.iasf.cnr.it}
             }

   \date{Received / Accepted}

   \abstract{
   We present the XMM-Newton observation of GRB 040106. From the X-ray spectral index and temporal decay, we argue that the afterglow is consistent with a fireball expanding in a wind environment. A constant density environment is excluded by the data. This is one of the very few cases in which this conclusion can be drawn. %Using the X-ray data, we have constrained the cooling frequency to be above the X-ray band at the start of the observation, $\sim$ 6 hours after the burst. Taking into account the optical observations, we have derived shalow constraints on the density of the surrounding medium, and the micro physic parameters.
   }
\authorrunning{Gendre et al.}
\titlerunning{XMM-Newton observation of GRB 040106}
   \maketitle
%
%________________________________________________________________

\section{Introduction}
Since the BeppoSAX revolution on the Gamma-Ray Burst (GRB) studies and the discovery of the GRB afterglows \citep{cos97}, a canonical model has emerged to explain the afterglow properties : the fireball model \citep{ree92, mes97, pan98}. This model is based on a blast wave which propagates into a surrounding medium, first considered to be uniform \citep[a review of this case is done by ][]{pir99}. But this model (referred as the InterStellar Medium [ISM] model) was not able to explain all the features in the afterglow spectra and light curves. First, some afterglow light curves displayed an achromatic break \citep[e.g.][]{pia01}. This was interpreted by the non isotropy of the blast wave , and this refined model was called the Jet model \citep{rho97, sar99}. Second, the optical afterglow light curves showed in some case a bump, associated with type Ic supernova \citep{rei99}. These and X-ray features \citep[e.g. ][]{ree02, piro99} show that long GRBs may be linked with hypernovae and star forming region \citep{mes01}. The density of the surrounding medium then decreases with the square of the distance to the central engine, due to the wind arising from the GRB progenitor \citep{che00}. This model is referred as the Wind model \citep{dai98, mes98, pan98, che99}.

The use of the multi-wavelength observations of a GRB afterglow allows one to determine the model parameters and to indicate the best model for each burst (and possibly to indicate why some bursts are best fitted with an ISM model while others are favoring a wind model). In this Letter, we will do so for GRB 040106, and we will show that the afterglow data of this burst indicate a fireball interacting in a wind. We will discuss this fact and the implications.

\section{GRB040106}
\object{GRB040106} was detected with the INTEGRAL Burst Alert System on 2004 January 6 at 17:55:12 UTC \citep{mer04}. It was detected in the IBIS/ISGRI data as a $\sim$ 60 seconds event. The peak flux was $6.5 \times 10^{-8}$ erg cm$^{-2}$ s$^{-1}$ in the 20-200 keV band \citep{got04}. A 45 kiloseconds long observation with XMM-Newton began at 23:11:23 UTC on 2004 January 6. The EPIC instruments were operating in full frame mode, with THIN filters (PN and MOS2) or MEDIUM filter (MOS1). One fading source was detected within the error box of the INTEGRAL detection \citep{ehl04}, and was associated with the X-ray afterglow of GRB040106. Using a cross-correlation with USNO-A2.0 stars, the refined position of this source is RA: 11 52 12.43, Dec: $-$46 47 15.9 \citep[J2000.0, 1 $\sigma$ total positional error 0.7'', ][]{ted04}.

The galactic position of this source is l = 292.5 and b = 14.88. In that direction, the column of density is N$_H =8.6 \times 10^{20}$ cm$^{-2}$ \citep{dic90}. The galactic optical extinction is E(B-V) = 0.1 \citep{sch98}. Optical observations detected an afterglow to this GRB, with an R magnitude of 22.4 $\pm$ 0.1 on 2004 January 7 at 08:33 and 23.7 $\pm$ 0.3 on 2004 January 8 at 08:25 \citep{mas04}. This imply a R magnitude corrected for the absorption of our galaxy of 22.1 $\pm$ 0.2 and 23.4 $\pm$ 0.4 (we assume a conservative 0.1 mag error in the reddening value).

A radio observation detected a source not consistent with the position of the X-ray afterglow \citep{wie04, ted04}. A later radio observation (on 2004 January 21.48 UT) with the VLA gave two upper limits of 100 $\mu$Jy and 170 $\mu$Jy at 8.46 and 4.86 GHz respectively \citep[2 $\sigma$ upper limits,][]{fra04}.

\section{Data analysis}

  \begin{table*}
      \caption[]{Spectral parameters of GRB040106. Each segment is indicated with the corresponding time range (in second) from the burst in the observer frame}
         \label{table_spectre}
     $$
         \begin{tabular}{cccccc}
            \hline
            \noalign{\smallskip}
            Spectrum      &   N$_H$ (10$^{20}$ cm$^{-2}$) & Energy spectral index & Flux (10$^{-12}$ erg s$^{-1}$ cm$^{-2}$) & $\chi^2_\nu$ & d.o.f.\\
            \noalign{\smallskip}
            \hline
            \noalign{\smallskip}
Complete                       & $8.8 \pm 0.8$ & 0.49 $\pm$ 0.04  & $ 1.01 \pm  0.02$ & 1.07 & 515\\
1$^{st}$ segment (21185-27178) & $10  \pm 3$   & 0.5  $\pm$ 0.1   & $ 1.7  \pm  0.1 $ & 1.06 & 64\\
2$^{nd}$ segment (27178-41512) & $ 8  \pm 2$   & 0.43 $\pm$ 0.05  & $ 1.25 \pm 0.02 $ & 1.16& 271\\
3$^{rd}$ segment (41512-63019) & $ 8  \pm 2$   & 0.5  $\pm$ 0.05  & $ 0.64 \pm  0.02$ & 0.99& 227\\
            \noalign{\smallskip}
            \hline
         \end{tabular}
     $$
   \end{table*}

We retrieved the Observation Data Files on the web page of the XMM-SOC, and processed them with the SAS 6.0 and the latest calibration files available. We have reduced the data using the {\it epchain} and {\it emchain} task of the SAS. Due to high background activity, we discarded 8 kiloseconds of observation, and kept only 37 kiloseconds of data. We have finally filtered the event files for good events (FLAG == 0 in the PN files, \#XMMEA\_EM for the MOS ones) and single or double patterns. We have extracted light curves and spectra using these filtered files, with a circular extraction region, with radius of 45 arcseconds (MOS) or 25 arcsecond (PN, avoiding thus any CCD gap). Background light curves and spectra were extracted using a larger (4 times the surface of the source extraction region) circular region on the same CCD.

\begin{figure}
   \centering
   \includegraphics[width=9cm]{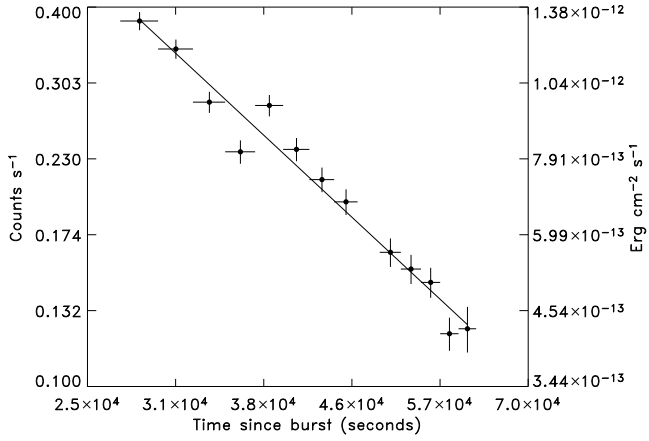}
      \caption{PN light curve of GRB040106 in the 0.2-12.0 keV band. The light curve is shown with the best fit power-law model. The flux indicated at the right of this figure is given in the 2.0-10.0 keV band.
              }
         \label{fig1}
   \end{figure}

Because this afterglow is bright (more than 10 000 photon detected), we decided to use the chi square statistic and binned the spectra in order to obtain at least 20 photons from the source (net photons) in each bin. We present the light curve in Fig. \ref{fig1}.

Fitting the light curves with a canonical power law, we obtain a decay index of $\delta$ = 1.4 $\pm$ 0.1 (we assume the notation F$_\nu \propto \nu^{-\alpha} t^{-\delta}$, all errors are quoted at the 90 \% confidence level). Such a decay is very common for a GRB afterglow.

\begin{figure}
   \centering
   \includegraphics[width=9cm]{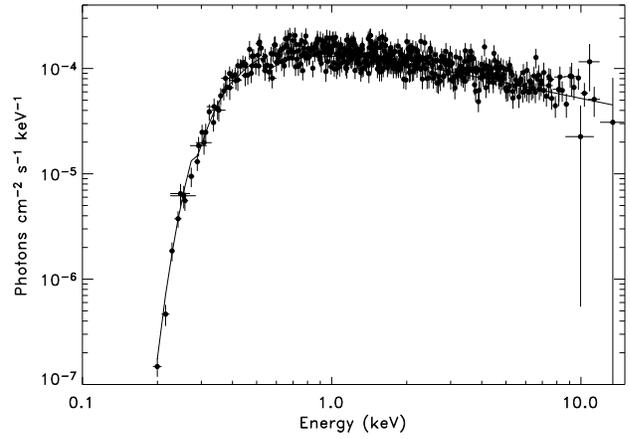}
      \caption{EPIC spectrum of GRB040106. The spectrum is shown with the best fit power law model.
              }
         \label{fig2}
   \end{figure}

We then made a spectral study of this afterglow. We first used an absorbed power law model on the whole exposure (presented in Fig. \ref{fig2}), fitting both EPIC-MOS and EPIC-PN spectra simultaneously, and allowing a normalization to vary between each instrument to take into account cross-calibration uncertainties. The best fit values are indicated in Table \ref{table_spectre}. We also looked for spectral variations. We divided the exposure in 3 segments with different durations (6, 12 and 19 kiloseconds of live time respectively), and repeated the spectral study. All the results are displayed in Table \ref{table_spectre}.

As one can see, the spectral parameter errors at the 90 \%
confidence level are compatible with no spectral variation.

In this paper, we are primarily interested in the properties of
the continuum. We thus carried out a simple analysis for line
detection. We added to the power law model an individual gaussian
line with a narrow width and free energy. We searched in the
summed and the three slices of pn spectra. We did not find strong
statistical evidence of line features. The most significant
features are found at 0.64 keV ($I=(10\pm3)\times10^{-6}$ ph
cm$^{-2}$ s$^{-1}$), 0.86 keV ($I=(5\pm1.5)\times10^{-6}$ ph
cm$^{-2}$ s$^{-1}$) and 3.6 keV ($I=(3.8\pm1.3)\times10^{-6}$ ph
cm$^{-2}$ s$^{-1}$) with an confidence level of 99.6\%, 99.2\% and
98.7\% respectively, as derived from the F-test. \footnote{a more
precise assessment of the statistical significance as indicated by
Protassov et al. (2002) would go beyond the scope of this paper}

\section{Constraints on the fireball model parameters}
Using both the spectral and temporal properties of this afterglow, we can investigate the parameters of the fireball model. We first tried to discriminate the type of fireball model (Wind, ISM or Jet models), using the closure relationships of \citet{sar98} and \citet{che99}. We derived values of $\delta - 1.5 \alpha = 0.64 \pm 0.13$ and $\delta - 2.0 \alpha = 0.42 \pm 0.14$ (90 \% confidence level). The value of $\delta - 1.5 \alpha$ should be used for the ISM and Wind models, while the value of $\delta - 2.0 \alpha$ applies to the Jet case. We should expect values of $-$0.5 or 0 for the the ISM model and $-$0.5 or 0.5 for the wind model, depending on the location of the cooling frequency, $\nu_c$. We thus conclude that our values are compatible only with a wind model. In the jet model, we should expect values of 0 or 1, which are not compatible with our values.

The closure relationship for the wind model are $\delta - 1.5 \alpha = -0.5$ if $\nu_c < \nu_x$ and $\delta - 1.5 \alpha = 0.5$ if $\nu_c > \nu_x$. Thus, we got a compatibility only if the cooling frequency is above the X-ray band.

May the cooling frequency pass through the X-ray band during the observation ? The light curve should show a steepening at that moment, with a decay variation of 0.25. There is a deviation in the light curve at about 12 kiloseconds (in the XMM-Newton observation time), which degrades the fit ($\chi^2_\nu$ = 2.37). % $\chi^2_\nu$ = 0.84).
We have tried to fit a broken power law to this light curve. The fit is not improved, indicating that the deviations are likely due to shot-time scale variations rather than to an overall steepening in the power law decay index. Moreover, the spectral index should indicate a flattening with a variation of 0.5 when the cooling frequency passes through the X-ray band. We can rule out such a large variation of the spectral index. Finally, our spectral index and temporal decay values are compatible with a cooling frequency above the X-ray band in all the three segments. We thus conclude that we indeed observe an afterglow in the wind model with the cooling frequency above the X-ray band.

Using the theoretical model, we can now constrain some model parameters. The theoretical temporal and spectral slopes are $\delta = (3p-1)/4$ and $\alpha = (p-1)/2$ respectively. We obtain $p = 2.2 \pm 0.2$ and $p = 2.0 \pm 0.1$ respectively.

We can also use the time at which the cooling frequency passes through the X-ray band (using an upper limit) to constrain the surrounding density medium. According to \citet{che00}, the cooling time observed in the X-ray band is :

\begin{equation}
\label{eqn1} t_c \sim 11.5 \times 10^{9}
\left(\frac{1+z}{2}\right)^3\left(\frac{\epsilon_B}{0.1}\right)^3E_{52}^{-1}A_*^4
{\rm ~Days}
\end{equation}

The redshift of this burst is unknown, we thus assumed the common
median value of 1. Using the data from the prompt emission, we
obtain $E_{\gamma, 52} \sim 1$. Because the cooling frequency is
above the X-ray even in the first part of the observation, $t_c <
0.23$ days. We thus obtain :

\begin{equation}
\label{eqn2} A_* < 2.0 \times 10^{-3}
\left(\frac{\epsilon_B}{0.1}\right)^{-3/4}E_{52}^{1/4}
\end{equation}

%By definition, the value of $\epsilon_B$ is less than 1. Thus, Eqn. \ref{eqn2} implies that $A_* < 1.2 \times 10^{-2}$. Using published values of $\epsilon_B$ from other burst \citep[$\epsilon_B \sim 10^{-4} - 10^{-2}$][]{pan02}, we obtain  $A_* \sim 10^{-5} - 10^{-3}$.

  \begin{figure}
   \centering
   \includegraphics[width=9cm]{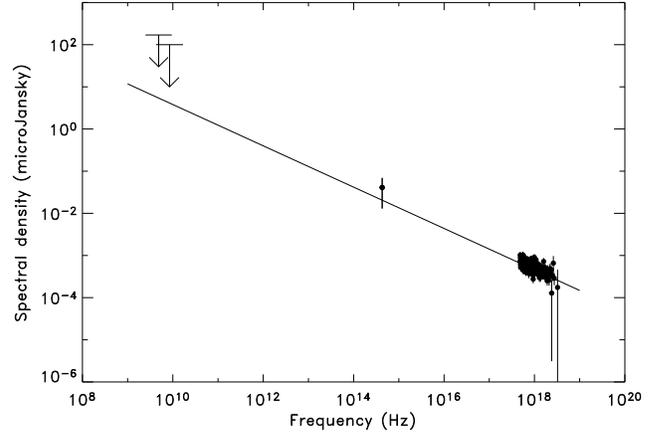}
      \caption{Spectrum of GRB040106. The spectrum is shown at 14.12 days after the burst. The power law indicated is the best fitted power law model from the X-ray data.
              }
         \label{fig3}
   \end{figure}

We now use the broadband spectrum between the radio and the X-ray band. The optical decay is 1.2 $\pm$ 0.4, fully consistent with the X-ray decay. Also, the unabsorbed optical-to-X-ray spectrum is compatible with a single power law, independently supporting the conclusion that $\nu_c$ is above the X-ray band. In addition, this indicates that $\nu_m$ is below the optical band at the date of the first optical observation (0.61 days). We use the expression of $\nu_m$ given by \citet{che00} :

\begin{equation}
\label{eqn3} \nu_m \sim 1.0 \times 10^{13}
\left(\frac{\epsilon_e}{0.1}\right)^{2}\left(\frac{\epsilon_B}{0.1}\right)^{1/2}E_{52}^{1/2}
{\rm ~Hz}
\end{equation}

Imposing $\nu_m < 4.28 \times 10^{14}$ Hz, we obtain :

\begin{equation}
\label{eqn4} \epsilon_e < 0.7
\left(\frac{\epsilon_B}{0.1}\right)^{-1/4}E_{52}^{-1/4}
\end{equation}

This condition does not give a very strong constraint on the value of $\epsilon_e$ : if $\epsilon_B$ is equal to one, then $\epsilon_e$ is less than $\sim$ 0.4.

Another constraint can be set by the flux density value in the X-ray. Using the equations given in \citet{pan00}, we got at 3 keV (where absorption is negligible) and at the time of the XMM-Newton observation :

\begin{equation}
\label{eqn5} F_\nu = 2.0 A_*
\left(\frac{\epsilon_e}{0.1}\right)^{1.1}\left(\frac{\epsilon_B}{0.1}\right)^{0.775}E_{52}^{0.775}
~\mu{\rm Jy}
\end{equation}

The flux density measured by XMM-Newton is $6.56 \times 10^{-2} \mu$Jy. We thus obtain :

\begin{equation}
\label{eqn6} A_* =  3.3 \times 10^{-2}
\left(\frac{\epsilon_e}{0.1}\right)^{-1.1}
\left(\frac{\epsilon_B}{0.1}\right)^{-0.775}E_{52}^{-0.775}
\end{equation}

 We have also verified if the position of $\nu_m$ can be constrained by using radio data. We have extrapolated the X-ray and optical fluxes at the date of the radio observation, assuming that there is no jet, and thus no achromatic break in the light curves. As one can see in Fig \ref{fig3}, the radio upper limits are compatible with the optical/X-ray spectrum. The presence of a jet would produce a steepening in the light curves \citep{rho97}, and thus lower optical/X-ray fluxes. This would also give a compatibility between the optical/X-ray fluxes and radio upper limits, preventing us to derive other constraints on $\nu_m$.

\section{discussion and conclusion}
From the X-ray observations of the afterglow of GRB 040106, we
have constrained the surrounding medium of the fireball to be a
wind environment. We have derived a constraint on the density of
the medium, which should be $A_* < 2 \times 10^{-3}
(\epsilon_B/0.1)^{-0.75}E_{52}^{1/4}$. By definition, if there is
no radiative losses, then $E_{52} = E_{\gamma,
52}/\epsilon_{\gamma}$, where $\epsilon_{\gamma}$ is the
conversion efficiency factor of the energy of the fireball into
gamma rays. We adopt here the value of $\epsilon_{\gamma} \sim
0.2$ as in \citet{fra01}.

Several GRB afterglows have been observed to be possibly in the
wind model \citep{che04}, although only in a very few cases the
wind profile is the only acceptable model (e.g. GRB011121). In
most of the cases, the $A_*$ value derived from the observations
is $\sim 0.3-0.7$, compatible with the Wolf Rayet star wind
observed in the Galaxy \citep{che99, che04}. But the cases of GRB
020405 \citep[$A_* \lesssim 0.07$][]{che04}, GRB 021211
\citep[$A_* = 0.0005$][]{kum03,che04} and GRB 011121 \citep[$A_* =
0.003$][]{pri02,pir04} indicate a value of $A_*$ significantly
lower than one. From present data, and using spectral modeling, we
can not significantly constrain $A_*$. If we set $\epsilon_B =
0.1$, $\epsilon_e = 0.3$, then $A_* = 2.8 \times 10^{-3}$. On the
other hand, values of $\epsilon_B = 10^{-4}$ \citep[observed by ][
in some afterglows]{pan02} and $\epsilon_e = 0.3$ imply that $A_*
= 0.53$.

\citet{che04} propose several interpretation for the low density sometime observed : a lower metalicity or a lower mass of the Wolf Rayet progenitors of GRBs, a GRB occurring along the progenitor rotation axis, and an unusual population of Wolf Rayet stars responsible for some GRBs. This could be tested by observations of a large set of Wolf Rayet stars and the host galaxies of GRB afterglows surrounded by a wind. Other XMM-Newton afterglow observations could grow the sample of GRBs surrounded by a wind with a good known location, and thus allow one to test these hypothesis.

\begin{acknowledgements}
This letter is based on observations made with XMM-Newton, an ESA science mission with instruments and contributions funded by ESA Members and the US. B.G. acknowledge the support of the EU through the EU FPC5 RTN 'Gamma ray burst, an enigma and a tool'. The authors thank Roger Chevalier for his usefull report.

\end{acknowledgements}


\begin{thebibliography}{}
\bibitem[Chevalier \& Li (1999)]{che99} Chevalier, R.A., \& Li, Z.Y., 1999, \apj, 520, L29
\bibitem[Chevalier \& Li (2000)]{che00} Chevalier, R.A., \& Li, Z.Y., 2000, \apj, 536, 195
\bibitem[Chevalier et al. (2004)]{che04} Chevalier, R.A., Li Z.Y., \& Fransson, C., 2004, \apj, 606, 369
\bibitem[Costa et al. (1997)]{cos97} Costa, E., Frontera, F., Heise, J., Feroci, M., in 't Zand, J., et al., 1997, Nature, 387, 783
\bibitem[Dai \& Lu (1998)]{dai98} Dai, Z.G., \& Lu, T., 1998, \mnras, 298, 87
\bibitem[Dickey \& Lockman (1990)]{dic90} Dickey, J.M., Lockman, F.J., 1990, \araa 28, 215
\bibitem[Ehle et al. (2004)]{ehl04} Ehle, M., Gonzalez-Riestra, R., \& Gonzalez-Garcia, B., 2004, GCN notice \#2508
\bibitem[Frail et al. (2001)]{fra01} Frail, D.A., Kulkarni, S.R,
Sari, R., et al., 2001, \apj, 562, L55
\bibitem[Frail et al. (2004)]{fra04} Frail, D.A., Wieringa, M., \& Soderberg, A.M., 2004, GCN notice \#2521
\bibitem[Gotz et al. (2004)]{got04} Gotz, D., Mereghetti, S., et al., 2004, GCN notice \#2506
\bibitem[Kumar \& Panaitescu (2003)]{kum03} Kumar, P., \& Panaitescu, A., 2003, \mnras, 346, 905
\bibitem[Masetti et al. (2004)]{mas04} Masetti, N., Palazzi, E., Rol, E., Pian, E., \& Pompei, E., 2004, GCN notice \#2515
\bibitem[Mereghetti et al. (2004)]{mer04} Mereghetti, S., Gotz, D., Beck, M., Borkowski, J., et al., 2004, GCN notice \#2505
\bibitem[Meszaros \& Rees (1997)]{mes97} Meszaros, P., \& Rees, M.J., 1997, \apj, 476, 232
\bibitem[Meszaros et al. (1998)]{mes98} Meszaros, P., Rees, M.J., \& Wijers, R.A.M.J., 1998, \apj, 499, 301
\bibitem[Meszaros (2001)]{mes01} Meszaros, P., 2001, Science, 291, 79
\bibitem[Panaitescu et al. (1998)]{pan98} Panaitescu, A., Meszaros, P., \& Rees, M.J., 1998, \apj, 503, 314
\bibitem[Panaitescu \& Kumar (2000)]{pan00} Panaitescu, A., \& Kumar, P., 2000, \apj, 543, 66
\bibitem[Panaitescu \& Kumar (2002)]{pan02} Panaitescu, A., \& Kumar, P., 2002, \apj, 571, 779
\bibitem[Pian et al. (2001)]{pia01} Pian, E., Soffita, P., Alessi, A., Amati, L., Costa, E., et al. 2001, \aap, 372, 456
\bibitem[Piran (1999)]{pir99} Piran, T., 1999, Physics Reports, 314, 575
\bibitem[Piro et al. (1999)]{piro99} Piro, L., Costa, E., Feroci, M., Stratta, G., Frontera, F., et al., 1999, \aaps, 138, 431
\bibitem[Piro et al. (2004)]{pir04} Piro, L., et al., 2004, in preparation
\bibitem[Price et al. (2002)]{pri02} Price, P.A., Berger, E., Reichart, D.E., Kulkarni, S.R., Yost, S.A., et al., 2002, \apj, 572, L51
\bibitem[Protassov et al. (2002)]{pro02} Protassov, R., van Dick, D. A., Connors, A., Kashyap, V. L. \& Siemiginowska, A., 2002,\apj, 571, 545
\bibitem[Rees \& Meszaros (1992)]{ree92} Rees, M.J., \& Meszaros, P., 1992, \mnras, 258, 41
\bibitem[Reeves et al. (2002)]{ree02} Reeves, J.N., Watson, D., Osborne, J.P., Pounds, K.A., O'Brien, P.T., et al., 2002, Nature, 416, 512
\bibitem[Reichart (1999)]{rei99} Reichart, D.E., 1999, \apj, 521, L111
\bibitem[Rhoads (1997)]{rho97} Rhoads, J.E., 1997, \apj, 487, L1
\bibitem[Sari et al. (1998)]{sar98} Sari, R., Piran, T., \& Narayan, R., 1998, \apj, 497, L17
\bibitem[Sari et al. (1999)]{sar99} Sari, R., Piran, T., \& Halpern, J.P., 1999, \apj, 519, L17
\bibitem[Schlegel et al. (1998)]{sch98} Schlegel, D.J., Finkbeiner, D.P., \& Davis, M., 1998, \apj, 500, 525
\bibitem[Tedds \& Watson (2004)]{ted04} Tedds, J.A., \& Watson, D., 2004, GCN notice \#2520
\bibitem[Wieringa \& Frail (2004)]{wie04} Wieringa, M., \& Frail, D.A., 2004, GCN notice \#2516


\end{thebibliography}
\end{document}